\documentclass[twocolumn,secnumarabic,amssymb, nobibnotes, superscriptaddress, nofootinbib,aps,prd]{revtex4-2}
\usepackage[colorlinks,citecolor=blue,linkcolor=blue,anchorcolor=blue,filecolor=blue, urlcolor=blue]{hyperref}
\usepackage{natbib}
\usepackage{xcolor}

\usepackage{amsmath}
\usepackage{adjustbox}
\usepackage{appendix}
\usepackage{graphicx}
\usepackage{booktabs}       
\usepackage{amsfonts}       
\usepackage{nicefrac}       
\usepackage{microtype}
\usepackage[capitalise]{cleveref}
\usepackage{xcolor}
\usepackage{multirow}
\usepackage{makecell}
\usepackage{comment}
\usepackage{todonotes}
\usepackage{array}
\usepackage{orcidlink}
\usepackage{float}
\usepackage{enumerate}
\usepackage{dsfont}
\usepackage{slashed}
\usepackage{siunitx}
\DeclareSIUnit{\Msun}{\ensuremath{M_{\odot}}}
\DeclareSIUnit{\nsat}{n_{\mathrm{0}}}
\DeclareSIUnit{\kb}{k_{\mathrm{B}}}
\DeclareSIUnit{\fm}{fm}
\usepackage[normalem]{ulem}

\newcommand{\diag}{\mathrm{diag}}

\begin{document}

\title{The petit four of color-superconducting phases in proto-neutron star evolution} 

\author{Selina Kunkel~\orcidlink{0009-0003-2942-0818}}
\email{kunkel@astro.uni-frankfurt.de}
\affiliation{Institut f\"{u}r Theoretische Physik, Goethe Universit\"{a}t, 
Max-von-Laue-Str.~1, D-60438 Frankfurt am Main, Germany}

\author{Ishfaq Ahmad Rather~\orcidlink{0000-0001-5930-7179}}
\email{rather@astro.uni-frankfurt.de}
\affiliation{Institut f\"{u}r Theoretische Physik, Goethe Universit\"{a}t, 
Max-von-Laue-Str.~1, D-60438 Frankfurt am Main, Germany}

\author{Hosein Gholami~\orcidlink{0009-0003-3194-926X}}
\email{mohammadhossein.gholami@tu-darmstadt.de}
\affiliation{Technische Universit\"{a}t Darmstadt, Fachbereich Physik, Institut f\"{u}r Kernphysik, Theoriezentrum, Schlossgartenstr.~2, D-64289 Darmstadt, Germany}

\author{Marco~Hofmann~\orcidlink{0000-0002-4947-1693}}
\email{marco.hofmann.physics@protonmail.com}
\affiliation{Technische Universit\"{a}t Darmstadt, Fachbereich Physik, Institut f\"{u}r Kernphysik, Theoriezentrum, Schlossgartenstr.~2, D-64289 Darmstadt, Germany}

\author{Jürgen Schaffner-Bielich~\orcidlink{0000-0002-0079-6841}}
\email{schaffner@astro.uni-frankfurt.de}
\affiliation{Institut f\"{u}r Theoretische Physik, Goethe Universit\"{a}t, 
Max-von-Laue-Str.~1, D-60438 Frankfurt am Main, Germany}

\newcommand{\MHcomment}[1]{\textcolor{blue}{[MH: #1]}}
\newcommand{\MHinsert}[1]{\textcolor{blue}{#1}}
\newcommand{\SK}[1]{\textsf{\color{red}{\textsuperscript{SK}#1}}}
\newcommand{\HG}[1]{\textsf{\color{purple}{\textsuperscript{HG}#1}}}
\newcommand{\IR}[1]{\textsf{\color{magenta}{\textsuperscript{IR}#1}}}
\newcommand{\JSB}[1]{\textsf{\color{violet}{\textsuperscript{JSB}#1}}}

\newcommand{\correct}[1]{\textsf{\color{red}{\sout{#1}}}}

\begin{abstract}

At high densities and moderate temperatures, hadronic matter is expected to undergo a first-order phase transition into a color-superconducting (CSC) state. 
A proto–neutron star describes the earliest evolutionary stages during the first seconds to minutes after core-collapse supernovae and therefore has the potential to assess the appearance of CSC phases at such high densities and moderate temperatures. 
To address this, we incorporate proto–neutron star conditions, considering neutrino-trapped and neutrino-transparent ones, into the equation of state including color-superconducting phases in a recently developed RG-consistent NJL-model. 
Since the total baryon number of a proto-neutron star is conserved during its later evolution, tracking stellar configurations from an initial mass of the hot proto–neutron star to the final cold neutron star along isolines of baryon number allows us to investigate whether color-superconducting phases can form at any point along this trajectory. 
By mapping this multidimensional transition in the hot furnace of a core-collapse supernovae cooling from a neutrino-trapped birth state to a cold, neutrino-transparent final state, we reveal four distinct core evolution scenarios--our "petit four" of proto–neutron star evolution: a delayed collapse from the CSC phase to a black hole, a persistent CSC phase, a vanishing CSC phase, and a fleeting CSC phase. 
For our specific parameterization of the hadronic and the CSC equation of state, we find that a stable color-superconducting phase can only be sustained in the final cold neutron star for a narrow, high-mass region.

\end{abstract}

\maketitle

\section{Introduction}
The study of neutron stars has a long history that started with the discovery of the neutron by Chadwick \cite{Chadwick:1932wcf} and shortly after the proposal of compact objects emerging from a supernova that mainly consist of nuclear matter by Baade and Zwicky \cite{Baade:1934wuu}. That these objects actually exist was later confirmed in 1967 with the discovery of the first four pulsars \cite{Hewish:1968bj}. In a small nod to the after-dinner speech of Jocelyn Bell Burnell \cite{1977NYASA.302..685B}, which has the title "Petit Four", we note that our analysis likewise identifies four distinct scenarios, in which color-superconducting phases appear, which are "baked" in the furnace of a core-collapse supernova during the birth of a proto-neutron star.

Neutron stars are the densest observable objects in the universe and provide valuable insight into matter at high densities, temperatures, and isospin asymmetries. They form when a massive star exhausts its nuclear fuel and explodes in a core collapse supernova \cite{Janka:2012wk, Bethe:1990mw}. In 1987, such an event was observed for the first time by measuring the neutrino flux emitted from a newly born proto-neutron star (PNS) \cite{Bionta:1987qt, Hirata:1988ad}. Early simulations of proto-neutron stars showed that in the first seconds to minutes following the explosion, these objects remain hot and neutrino-rich before cooling via extensive neutrino emission \cite{Burrows:1986me}. Advanced numerical models have since confirmed this evolutionary picture and demonstrated the critical impact of convection and neutrino transport on the macroscopic properties of the star \cite{Keil:1996ab, Pascal:2022qeg, 1988PhR...163...51B, 1996ApJ...473L.111K, Pons:2001ar, Fischer:2020xjl, Jakobus:2022ucs, Fischer:2010wp, Fischer:2017lag}. 

Similar to neutron stars at zero temperature, the exact nature of the matter inside the core of a proto-neutron star remains an open question. Various forms of exotic matter have been considered in previous works and implemented into the equation of state to study mass-radius configurations and the long-term evolution of proto-neutron stars \cite{Glendenning:1992vb, Alcock:1986hz, Bombaci:2004mt, Weber:2004kj, Buballa:2003qv, Sagert:2008ka, Malfatti:2019tpg, Sabatucci:2026qcz, Roark:2018boj, Roark:2018uls}. 

In this work, we focus on the effects of color superconductivity (CSC) in proto-neutron stars, systematically tracking which phases the core can access along its evolutionary path. The underlying microscopic mechanism generating color superconductivity is based on the Bardeen-Cooper-Schrieffer (BCS) theory, which states that fermions can form pairs if an attractive interaction exists near the Fermi surface \cite{Bardeen:1957mv}. In quantum chromodynamics (QCD), a similar pairing mechanism occurs for quarks interacting via the strong interaction, leading to color superconductivity \cite{Barrois:1977, Bailin:1984}. This state of matter exhibits a rich phase structure due to additional flavor and color degrees of freedom, with the most prominent configurations being the two-flavor superconductor (2SC) and the color-flavor-locked (CFL) phase \cite{Alford:1998mk}. Consequently, color superconductivity is a well-established candidate for the high-density cores of compact objects and has been a hot topic of research for the past several years \cite{Alford:2007xm, Ruester:2003zh, Baldo:2002ju, Ruester:2005jc, Sandin:2007zr}. Furthermore, recent studies~\cite{Roupas:2020nua, Gholami:2024ety, Geissel:2025vnp}, including constraints on twin star configurations~\cite{Christian:2025dhe}, have been widely used to refine these model parameters. Beyond cold matter, several investigations have explored color-superconducting quark matter under explicit proto-neutron star conditions, including finite temperature and neutrino trapping \cite{Ruester:2005ib, doCarmo:2013btr, Alford:2025jtm, Gholami:2025yqf, Sabatucci:2026qcz}.

These works demonstrate that stable color-superconducting regions can exist in both hot and cold stellar environments. But because a cold neutron star is the direct end-product of a hot, lepton-rich proto-neutron star, it is vital to investigate whether a cooling track can actually reach the stable color-superconducting branch of the cold mass-radius curve. Therefore, the main objective of this work is to determine if color superconductivity can be continuously realized during the early stages of evolution, or if the core undergoes full reversion back to hadronic matter as it cools. We emphasize that these structural trajectories are model-dependent; the accessibility of specific paths relies heavily on the chosen dense-matter framework, the precise onset density of the phase transition, and the initial maximum-mass configuration at birth.

The outline of the paper is as follows: We will first present an introduction to proto-neutron star evolution and then present the RG-consistent NJL model used to model color-superconducting matter. The distinction between neutrino-transparent and neutrino-trapped beta-equilibrium will be explained afterwards. The theoretical part will conclude with a chapter on how the mixed phase between hadronic and quark matter is constructed. Then we will present phase diagrams including isentropes in the neutrino-transparent and neutrino-trapped case. We will conclude with mass-radius diagrams and show the evolutionary trajectory of the proto-neutron star for conserved particle number.

\section{Methodology}\label{NS1}
\subsection{Proto-Neutron Star Evolution}

\begin{figure*}[t]
\includegraphics[width=0.9\textwidth]{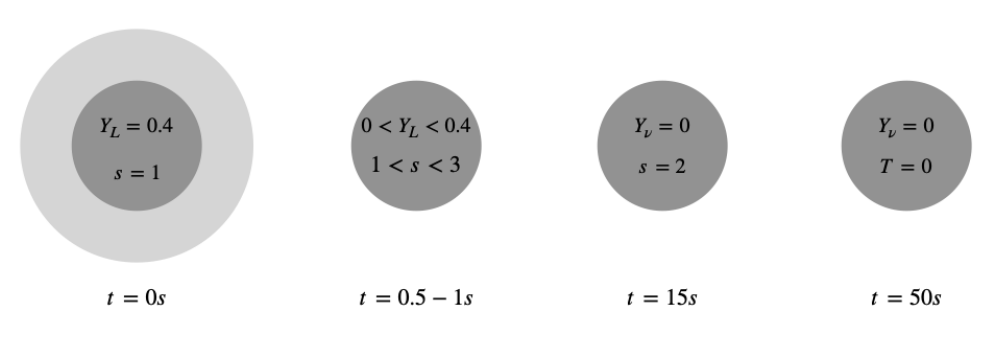}
\caption{Schematic overview of proto-neutron star evolution. The stages trace the long-term transition from an early neutrino-trapped, high-entropy birth configuration to a neutrino-transparent, cold isolated remnant. The condition $Y_{\nu} = 0$ denotes a neutrino-transparent state where the net neutrino number (the difference between neutrino and anti-neutrino densities) vanishes.}
\label{fig:pns_evolution}
\end{figure*}

A neutron star is born in a so-called core-collapse supernova, which occurs after a massive progenitor star exhausts its nuclear fuel and forms an iron core. Because iron is the most stable nucleus, further exothermic fusion ceases, causing the core to lose the thermal pressure required to support itself against gravitational collapse. During the resulting contraction, photodisintegration processes absorb thermal energy from the system, while rapid electron capture (inverse beta decay) reduces electron degeneracy pressure, driving the core into free-fall collapse. The collapse abruptly halts once the central core reaches nuclear saturation density, where the short-range repulsive component of the nuclear force overrules gravity. The inner core stiffens, causing the infalling outer matter to bounce and launch an energetic outgoing shock wave. The central remnant left behind is a hot, lepton-rich proto-neutron star that eventually evolves into a cold, catalyzed configuration.

The thermodynamic and compositional evolution of a proto-neutron star proceeds through several well-defined epochs~\cite{Prakash:1996xs, Pons:1998mm, Kunkel:2024otq}. Detailed dynamical simulations showed that the radial profile of the proto-neutron star is nearly isentropic throughout evolution, justifying the use of isentropic equations of state in our calculations. The evolution is studied under the assumption that the hydrodynamic timescale is much shorter than the Kelvin-Helmholtz cooling timescale. It is a quasi-static approach in which the stellar structure is obtained from hydrostatic equilibrium at each timestep. The entropy and the lepton number are governed by neutrino transport. 

The evolution proceeds as follows: Immediately, after core bounce, the lepton fraction in the PNS equals approximately the one which was present in the degenerate iron core before, initially at approximately $Y_L = n_L/n_B \approx 0.4$. At this stage, the extremely high interior temperatures result in a short neutrino mean-free path, trapping neutrinos within the stellar interior. The initial entropy per baryon in the inner core is relatively low, around $s = 1$ (entropy per baryon in units of $k_B$), while the unshocked outer envelope can reach values of $s = 5 - 10$ due to ongoing matter accretion. Within approximately $0.5\,$seconds, the shock wave detaches from the envelope to drive the supernova explosion, initiating a phase of rapid deleptonization as trapped neutrinos diffuse out of the star. As these neutrinos escape, their diffusion converts chemical energy into thermal energy; this process, combined with gravitational contraction, heats the inner core to a maximum entropy of $s \approx 2$. After $10 - 15\,$seconds of continuous diffusion, the star becomes entirely transparent to neutrinos, signaling that weak interactions have reached full equilibrium \cite{Janka:2012wk}. 

Throughout the first 15 seconds after core bounce, structural modifications can potentially trigger a delayed gravitational collapse to a black hole. If the early envelope accretes a critical amount of fallback material, the total stellar mass may exceed the maximum stable limit supported by the hot equation of state. Alternatively, the progressive deleptonization of the stellar core modifies the composition and equation of state, lowering the threshold for the onset of deconfined quark matter or strange hyperons. This localized softening of the equation of state reduces the maximum supportable mass, inducing an instability. Over the subsequent 50 seconds, neutrino cooling drives the remnant down toward its cold, catalyzed configuration. While the stellar crust remains relatively warm initially, complete thermal equilibrium throughout the entire star is achieved over the next century, primarily mediated by weak Urca cooling processes. A structural timeline of this multi-stage evolutionary sequence is illustrated schematically in \cref{fig:pns_evolution}.

\subsection{RG-consistent NJL model for the quark phases}
We describe deconfined quark matter within a three-flavor Nambu--Jona--Lasinio (NJL) model that incorporates both dynamical chiral symmetry breaking and color superconductivity. The quark sector is defined by the Lagrangian \cite{Rehberg:1995kh,Gastineau:2001zke,Klahn:2006iw}

\begin{align}
\label{eq:Lagrangian}
\mathcal{L}&=\bar{\psi}(i\slashed{\partial}+\gamma^0\hat{\mu}-\hat{m})\psi \nonumber\\
&+G_S\sum_{A=0}^8\left[(\bar{\psi}\tau_A\psi)^2 + (\bar{\psi} i \gamma_5 \tau_A \psi)^2\right]
\nonumber\\
&- K \bigl[\det_{\text{f}}\bigl(\bar{\psi}(\mathds{1}+\gamma_5)\psi\bigr) + \det_{\text{f}}\bigl(\bar{\psi}(\mathds{1}-\gamma_5)\psi\bigr)\bigr] \nonumber\\[1ex]
&+ G_D
\big[\,\big(\bar{\psi}^a_\alpha i\gamma_5 \epsilon^{\alpha\beta\gamma} \epsilon_{abc}(\psi_C)^b_\beta\big)
\big((\bar{\psi}_C)_r^\rho i\gamma_5 \epsilon_{\rho\sigma\gamma} \epsilon^{rsc} \psi_s^\sigma\big)\nonumber\\
&\hspace{3em}+\; \big(\bar{\psi}^a_\alpha  \epsilon^{\alpha\beta\gamma} \epsilon_{abc}(\psi_C)^b_\beta\big)
\big((\bar{\psi}_C)_r^\rho  \epsilon_{\rho\sigma\gamma} \epsilon^{rsc} \psi_s^\sigma\big) \,\big]\nonumber \\
&-G_V\bigl(\bar\psi\gamma^{\mu}\psi\bigr)^2+\mathcal{L}_{\text{lep}}\ 
\end{align}

where $\psi$ denotes the quark field carrying flavor ($u,d,s$) and color ($r,g,b$) degrees of freedom. The matrices $\hat m$ and $\hat \mu$ represent the quark mass and chemical potential matrices, respectively, while $\tau_A$ are the Gell-Mann matrices in flavor space supplemented by $\tau_0=\sqrt{2/3}\,\mathds{1}_f$. Charge-conjugated spinors are defined as $\psi_C=C\bar\psi^T$ and $\bar\psi_C=\psi^T C$, with $C=i\gamma^2\gamma^0$.

The interaction terms encode the essential low-energy features of QCD. The scalar--pseudoscalar four-fermion interaction governed by $G_S$ generates spontaneous chiral symmetry breaking, and the determinantal six-quark interaction proportional to $K$ represents the instanton effects which break the axial $U(1)_A$ symmetry \cite{Kobayashi:1970ji,tHooft:1976rip}. Attractive quark pairing is introduced through the coupling $G_D$, which acts in the color- and flavor-antitriplet channel and allows for the formation of color-superconducting condensates. A repulsive vector interaction controlled by $G_V$ is included to stiffen the quark equation of state and thereby support hybrid stars compatible with the observed two-solar-mass constraint \cite{Klahn:2006iw}.

Throughout this work, the model is treated in the mean-field approximation. Diquark pairing is characterized by the gap parameters

\begin{align}
\Delta_{c} = 2 G_D\langle(\bar{\psi}_C)_{\alpha}^{a} i \gamma_5
\epsilon^{\alpha \beta c} \epsilon_{a b c} \psi_{\beta}^{b}\rangle  
\end{align}

where $\Delta_1$, $\Delta_2$ and $\Delta_3$ correspond to the $ds$, $us$ and $ud$ pairing channels. The constituent quark masses are given by 

\begin{equation}
M_\alpha = m_\alpha - 4 G_S \sigma_\alpha + 2 K \sigma_\beta \sigma_\gamma \; ,
\label{Mi}
\end{equation}

where the chiral condensates $\sigma_\alpha=\langle\bar\psi_\alpha^a\psi_\alpha^a\rangle$ determine the dynamically generated masses.

For fixed temperature $T$ and baryon chemical potential $\mu_B$, the equilibrium state is obtained by minimizing the mean-field effective potential $\Omega_{\rm eff}$. The chiral, diquark, and vector condensates satisfy the corresponding gap equations, and among all solutions, we select the one with the lowest free energy. Electric and color neutrality are enforced simultaneously by extremizing $\Omega_{\rm eff}$ with respect to the associated chemical potentials.

Since the NJL model is non-renormalizable, we regularize it using a sharp three-momentum cutoff $\Lambda'=602.3\,\mathrm{MeV}$. The vacuum parameters are fitted to reproduce the meson spectrum \cite{Rehberg:1995kh}, while medium calculations are performed within the renormalization-group consistent framework of Ref.~\cite{Gholami:2024diy}, see also Ref.~\cite{Braun:2018svj}, employing the RG-consistent massless scheme throughout this work. The equation of state can be reproduced with the open-source NJL
module \cite{hofmann_2026_18249033}, published as part of the most recent \textit{Calliope} version of the MUSES Calculation Engine~\cite{Jahan:2026hvs}. We use the parameter set $G_D=1.45G_S$ and $G_V=0.7G_S$, corresponding to parameter set 1 of Ref.~\cite{Gholami:2024ety}, which yields hybrid-star configurations consistent with current astrophysical constraints.

The leptonic sector consists of electrons, muons, and, when neutrino trapping is considered, their corresponding neutrinos, all treated as free relativistic Fermi gases. The thermodynamic state is specified by the conserved quark number, electric charge, color charges, and, in neutrino-trapped matter, the lepton number. The associated chemical potentials are introduced through the matrix

\begin{equation}
\mu_{ab}^{\alpha\beta} = \left(
  \mu \delta^{\alpha\beta}
+ \mu_Q Q^{\alpha\beta} \right)\delta_{ab}
+ \left[ \mu_3 \left(T_3\right)_{ab}
+ \mu_8 \left(T_8\right)_{ab} \right] \delta^{\alpha\beta}~,
\label{mu-f-i}
\end{equation}

where
\begin{align}
Q&=\diag_f\left(\frac23,-\frac13,-\frac13\right), \\
T_3&=\diag_c\left(\frac12,-\frac12,0\right), \\
\sqrt{3}\,T_8&=\diag_c\left(\frac12,\frac12,-1\right).
\end{align}
Neutrinos carry lepton number only,
\begin{align}
   \mu_{\nu_e} = \mu_{L_e}~, \qquad
   \mu_{\nu_\mu} = \mu_{L_\mu}~,
\end{align}
whereas charged leptons carry both lepton number and electric charge,
\begin{align}
   \mu_e  = \mu_{L_e} - \mu_Q~, \qquad
   \mu_\mu  = \mu_{L_\mu} - \mu_Q~.
\end{align}
The relations between the quark and lepton chemical potentials follow from the assumed beta-equilibrium condition.

\subsection{Relativistic mean-field model for the hadronic phase}

To describe the low-density hadronic phase of the NS interior, we employ the DD2 EoS~\cite{Typel:2009sy}. The DD2 model is a relativistic mean-field (RMF) framework based on a generalized relativistic density-dependent nucleon-meson field theory, where the coupling constants of the $\sigma$, $\omega$, and $\rho$ mesons to the nucleons are parameterized as explicit functions of the baryon density. This density dependence naturally incorporates many-body correlation effects without requiring non-linear meson self-interaction terms, providing an excellent description of the properties of finite nuclei and symmetric nuclear matter at saturation density~\cite{Typel:2009sy, Hempel:2009mc}.
Crucially for nuclear astrophysics, the DD2 EoS yields a symmetry energy slope parameter ($L \approx 55\text{ MeV}$) that aligns well with modern experimental constraints~\cite{Lattimer:2023rpe, Essick:2021kjb, Lim:2023dbk} and generates a maximum mass for cold, purely hadronic neutron stars well above $2\,M_\odot$~\cite{Typel:2009sy}. Due to its robust thermodynamic consistency across wide ranges of temperature, density, and proton fraction, the DD2 framework has been widely adopted in core-collapse supernova simulations and general-relativistic hydrodynamic models of binary neutron star mergers~\cite{Hempel:2009mc, Fischer:2013eka}.

\subsection{Neutrino-transparent versus neutrino-trapped beta equilibrium}\label{neutrino_chapter}

In this work, we consider two different scenarios in order to fix the charge chemical potential $\mu_Q$ and the chemical potentials of the leptons: neutrino-transparent beta equilibrium and neutrino-trapped beta equilibrium.

In the scenario of neutrino-transparent beta equilibrium we assume that the star is transparent to neutrinos, such that they leave the system. In this case, neutrinos do not contribute to the equation of state and lepton number is not conserved. In the hadronic phase, composed of $npe$ matter, detailed balance of the direct Urca processes
\begin{align}
    n &\to p+e^- + \bar{\nu}_e \text{ (neutron decay), and}\\
    p+e^- &\to n+ \nu_e \text{ (electron capture),}
\end{align}
impose the relation
\begin{equation}
    \mu_n = \mu_p +\mu_e.
\end{equation}
In the quark-matter phases, detailed balance of the processes
\begin{align}
	&d \to u + e^- + \bar{\nu}_e\,, \\
	&s \to u + e^{-} + \bar{\nu}_e\,, \\
	&u + e^{-} \to d + \nu_e\,, \\
	&u + e^{-} \to s + \nu_e\,,
\end{align}
impose
\begin{equation}
    \mu_d=\mu_s = \mu_u + \mu_e.
\end{equation}
Electric charge and baryon number are the only conserved charges, thus the chemical potentials for electrons and muons are identical and given by $\mu_e = \mu_\mu=-\mu_Q$, with $\mu_Q$ fixed by the electric charge neutrality.

In the scenario of neutrino-trapped beta equilibrium, we assume that the mean-free paths of the neutrinos are short enough compared to the system size, such that the neutrinos get trapped in the system and thermalize, i.e. they form an equilibrium (Fermi-Dirac) distribution. In this scenario, lepton number is conserved and the beta-equilibrium condition reads
\begin{equation}
    \mu_{\nu_e} + \mu_n = \mu_p + \mu_e
\end{equation}
in the hadronic phase, and
\begin{equation}
    \mu_d = \mu_s = \mu_u + \mu_e - \mu_{\nu_e}
\end{equation}
in the quark-matter phases, respectively. Here, the chemical potentials of the leptons are related to the electron lepton-number chemical potential $\mu_{L_e}$ as 
\begin{align}
    \mu_e &= \mu_{L_e} - \mu_Q, \label{eq:mu_e_trapped}\\
    \mu_{\nu_e}&=\mu_{L_e} \label{eq:mu_nu_trapped}.
\end{align}
For the hadronic phase, the DD2 EoS is already tabulated in the standardized CompOSE format \cite{CompOSECoreTeam:2022ddl,Typel:2013rza,Oertel:2016bki,CompOSE} and can be interpolated to fix a constant desired electron lepton-number fraction $Y_{L_e}$.
In order to calculate the quark-matter EoS at a constant electron $Y_{L_e}$ we proceed via the following steps:
\begin{enumerate}
    \item We first tabulate the quark-matter EoS as a three-dimensional table with $T,\mu_B,\mu_Q$ as free parameters.
    \item For a given $T,\mu_B,\mu_Q$ in the table, we then add electrons to the EoS. The electron chemical potential $\mu_e$ is fixed by requiring electric charge neutrality $n_Q=0$ of the system including quarks and electrons.
    \item Given $\mu_Q$ and $\mu_e$, \cref{eq:mu_e_trapped} now fixes the electron lepton-number chemical potential $\mu_{L_e}$, and equilibrated electron neutrinos are added to the system with this chemical potential $\mu_{\nu_e}=\mu_{L_e}$ (\cref{eq:mu_nu_trapped}).
    \item The three-dimensional EoS table of the system with quarks, electrons and electron-neutrinos with the free parameters $T,\mu_B,\mu_Q$ is now interpolated along a fixed electron lepton-number fraction $Y_{L_e}=\text{constant}$ in the plane spanned by $T,\mu_B$. From this two-dimensional table, one-dimensional slices at constant $T$ can be obtained or slices at constant entropy per baryon $s=\text{constant}$ can be interpolated for PNS evolution.
\end{enumerate}

Note that for the quark-matter EoS, we include muons in the neutrino-transparent beta-equilibrium EoS, but do not include muons or muon neutrinos in the neutrino-trapped beta-equilibrium EoS for simplicity. Alternatively, we could have added the muon neutrinos also in neutrino-trapped equilibrium. Neglecting neutrino oscillations, muons could have been added either assuming muon-neutrino trapping, giving rise to a conserved muon lepton-number $Y_{L_\mu}$, or muons could have been added without muon neutrinos, assuming free streaming of the muon neutrinos, i.e. $\mu_{\mu}=-\mu_Q$. For simplicity, we refrain from fixing a muon lepton fraction during PNS evolution, but describe the evolution only in terms of fixed $n_B,Y_{L_e}$ and $s$.
The propagation of neutrinos within the RG-NJL model has been studied in detail in  \cite{Alford:2025jtm}.

\subsection{Construction of a mixed phase}
When looking at proto-neutron stars one typically fixes the entropy per baryon so that one switches to an isentropic system. Unlike the isothermal case, where a transition occurs directly at equal pressure while maintaining constant temperature, the isentropic case is more restrictive: since the two phases typically have different entropies at the phase boundary, the system has to move along the coexistence line until entropy is matched. At a fixed temperature, the CFL phase typically has a smaller entropy per baryon because it has less degrees of freedom due to all quarks being gapped. 
In the neutrino-transparent case the phase transitions for a fixed temperature are done with a Maxwell construction, but the mixed phase of the isentropes has to be done either by interpolation or by constructing the needed quantities via volume fractions. 

For each temperature, we check whether the desired value for the entropy per baryon is in one of the two phases. If it lies directly between the phases we calculate the mixed phase in the following way: \footnote{We note that the entropy per baryon is not a thermodynamic potential and can strictly speaking not be used for the approach with volume fractions. However, the pressure and energy density are calculated via linear interpolation in the mixed phase, so that the TOV results are not affected by our choice.}

\begin{equation}
    s = \chi_{2SC} \cdot s_{2SC}+(1- \chi_{2SC}) \cdot s_{CFL}.
    \label{eq:mixed-phase-2}
\end{equation}

Because the volume fractions $\chi_i$ summed up have to result in 1, we use $\chi_{\text{2SC}} + \chi_{\text{CFL}} = 1 \Leftrightarrow \chi_{\text{CFL}}= 1 - \chi_{\text{2SC}}$. 

We take the entropy per baryon value of the two phases directly before and after the transition, insert them into \cref{eq:mixed-phase-2} and solve for $\chi_{2SC}$ for the isentrope in question.

Transitioning to a neutrino-trapped system leads to fixing a lepton fraction and an additional chemical potential, which is the lepton chemical potential. Usually one constructs the mixed phase in such a case with a Gibbs construction as it was done in \cite{Sabatucci:2026qcz}. A phase transition with a Gibbs construction occurs when the pressure of the two phases are the same at the same temperature, the same baryon chemical potential and the same charge chemical potential. For this situation the hadronic and the quark equation of state would need to be set up in dependence of the same quantities. The DD2 equation of state has the temperature, proton fraction and number density as independent quantites, while the quark equation of state has temperature, baryon chemical potential and charge chemical potential as independent quantities. 

Hence, we match the neutrino-trapped hadronic equation of state to the
neutrino-trapped color-superconducting equation of state with a Maxwell
construction at the same temperature, and the mixed phase is then constructed
via the approach explained above. This enforces local charge neutrality in
each phase and corresponds to the limit of large surface tension. A Gibbs
construction~\cite{Glendenning:1992vb}, valid for small surface tension, would
instead allow globally-neutral charged phases to coexist and broaden the mixed
phase, as employed in Ref.~\cite{Sabatucci:2026qcz}. Which of the two is
realized depends not only on the (poorly constrained) quark--hadron surface
tension but also on whether the corresponding charged structures can form at all: their nucleation competes with the dynamical
timescale of the phase conversion. As both the interface tension
and these formation timescales are essentially unknown, we adopt the Maxwell
construction as a well-defined limiting case.

Within the RG-NJL model, the matching between the 2SC and CFL phase is performed using a similar approach as a Gibbs construction. At each $T,\mu,\mu_Q$ slice we compare the pressure of the 2SC and CFL phase and take the phase with the higher pressure as the stable phase for this particular combination of temperature, quark chemical potential and charge chemical potential. Then we interpolate for fixed $Y_L$ via the approach explained in \cref{neutrino_chapter} and if the desired value for the lepton fraction lies inbetween a 2SC and CFL phase we interpolate linearly and declare this point as a mixed phase point.


\section{Results}
\label{results}

\subsection{Phase diagrams}

\begin{figure}[t]
\includegraphics[width=0.5\textwidth]{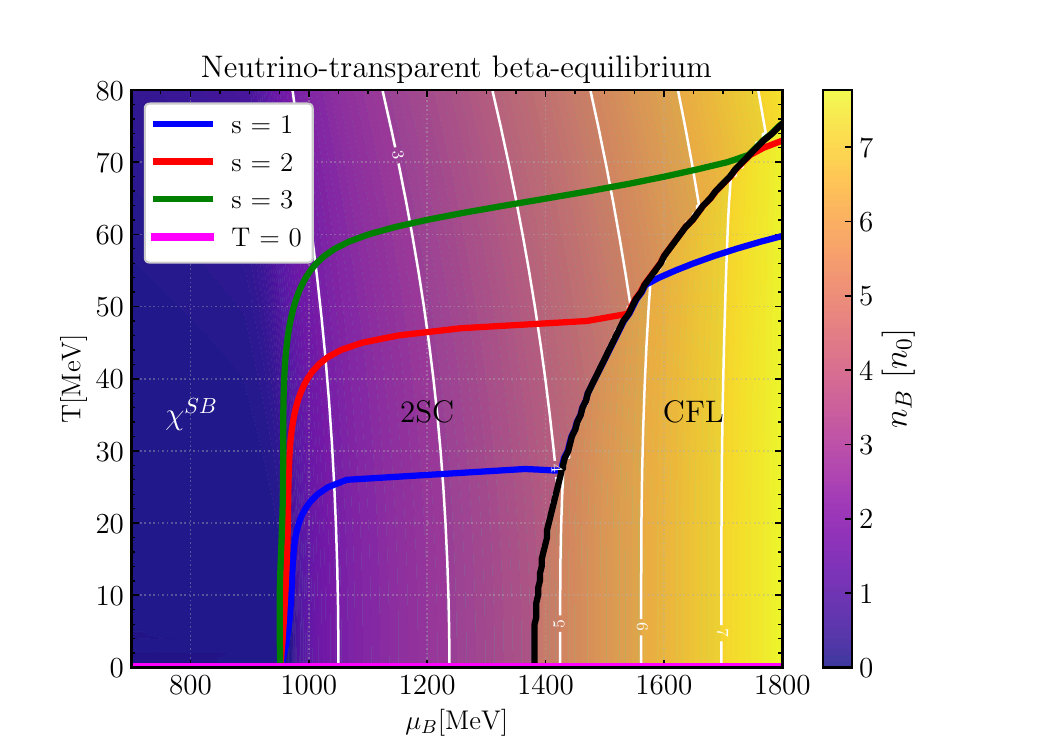}
\caption{Phase diagram of quark matter within the RG-NJL model in the T-$\mu_B$ plane in neutrino-transparent beta-equilibrium. The chiral broken, 2SC, and CFL phases are shown together with isentropic trajectories for different values of entropy per baryon, s=1,2,3, and the zero-temperature limit. The color map indicates the baryon number density with isodensity contours at integer multiples of the nuclear saturation density.}
\label{fig:pd_NJL1}
\end{figure}

In this section, the phase structure and the trajectories of the isentropic equations of state are explored. \cref{fig:pd_NJL1} shows the trajectories of the isentropes in the plane of temperature and baryon chemical potential in quark matter within the RG-NJL model for the neutrino-transparent beta-equilibrium case. The color coding represents the baryon number density with isodensity contours drawn at integer multiples of nuclear saturation density. At zero temperature, we observe a smooth transition from the chiral broken phase to the 2SC phase, which shifts to lower baryon chemical potentials for higher temperatures. Meanwhile, the transition between the 2SC and CFL phase is first order and shifts to higher baryon chemical potentials and densities for higher temperatures. 

    At zero temperature, the density jumps from $\SI{3.7}{\nsat}$ to $\SI{4.7}{\nsat}$ at the transition from 2SC to CFL. 
The blue, red, and green lines show the trajectories of the isentropes for values of $s = 1$, $s = 2$, and $s = 3$, respectively. All curves start at the same point at zero temperature and have a steep increase in temperature close to the $\chi_{\text{SB}}-$2SC phase boundary. At similar chemical potentials, the curves then start to form a plateau in the 2SC phase at different temperatures, which is most pronounced for $s = 1$. This is related to a large heat capacity coming from a high number of degrees of freedom, mainly being the unpaired blue quarks in the 2SC phase, so that while compressing the fluid the temperature does not change much. This microphysical feature directly underpins the distinct flat plateau features we talk about in the Mass-Radius section next \cref{sec:mr_chapter_transparent}.

The increase of the isentrope at the boundary between the 2SC and CFL phase is a result of the underlying microphysics.
The CFL phase is fully
gapped, so its entropy density $s$ is strongly suppressed: only quasiparticles
with ungapped modes near the Fermi surface can be thermally excited, and in
the 2SC phase these are the unpaired blue and strange quarks, whereas in CFL
all quark modes acquire a gap (see for instance \cite{Gholami:2025yqf}). The baryon density $n$, in contrast, is set by
the bulk of the filled Fermi sea and is therefore comparable in the two
phases (if anything slightly larger in the denser CFL phase). Both effects act
in the same direction, so the suppressed entropy density translates into a
smaller entropy per baryon $s/n$ in the CFL phase, and an isentrope must
correspondingly move to \emph{higher} temperature along the 2SC--CFL boundary,
consistent with \cref{fig:pd_NJL1}

\begin{figure}[t]
\includegraphics[width=0.5\textwidth]{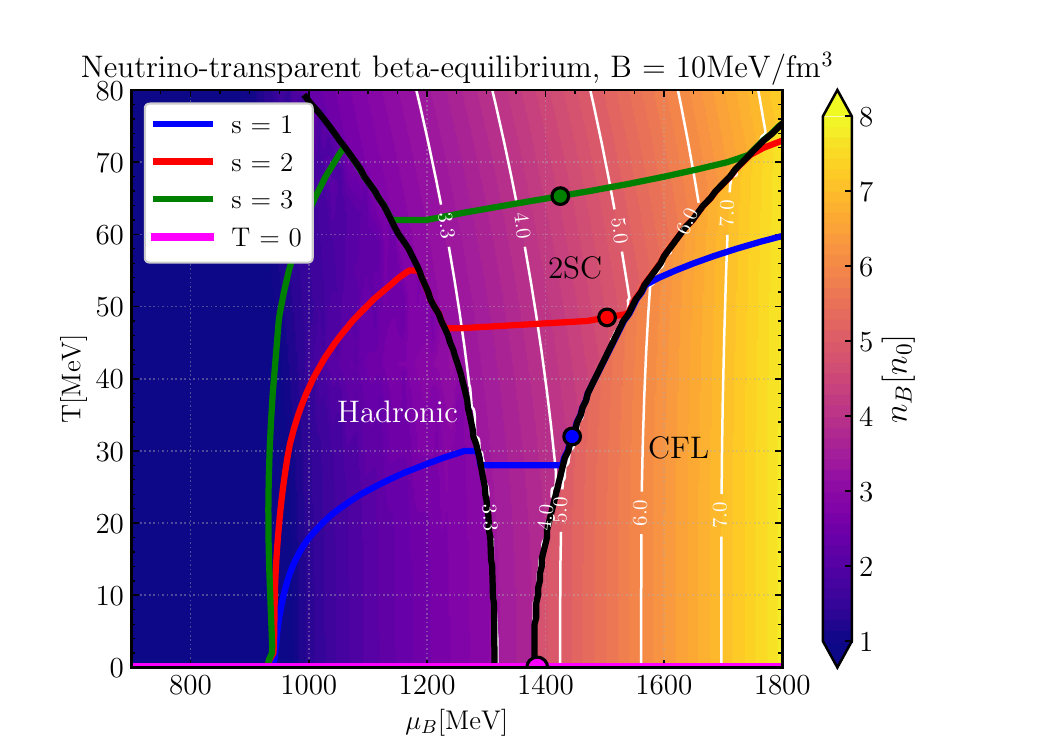}
\caption{Phase diagram of color-superconducting matter within the RG-NJL model matched to the hadronic DD2 equation of state at low densities in the T-$\mu_B$ plane in neutrino-transparent beta-equilibrium. The hadronic, 2SC, and CFL phases are shown together with isentropic trajectories for different values of entropy-per-baryon $s=1,2,3$ and the zero-temperature limit. The color map indicates the baryon number density with isodensity contours at integer multiples of the nuclear saturation density. The filled circles indicate matter in the center of the maximum-mass configuration of the hybrid EoS, see the corresponding mass-radius curves in \cref{fig:mr_yn0}.}
\label{fig:pd_DD2NJL1}
\end{figure}

\cref{fig:pd_DD2NJL1} represents the case where the DD2 EoS has been matched to the RG-NJL1 model at lower densities. The matching was done by the inclusion of an additional bag constant. With a nonvanishing bag constant the pressure and energy density change in the following way:

\begin{equation*}
P \rightarrow P - B, \qquad \epsilon \rightarrow \epsilon + B
\end{equation*}

The additional bag constant was chosen to be $B = \SI{10}{\MeV\per\fm^3}$. This value ensures that there is a 2SC phase at zero temperature and a matching to quark matter also at higher temperatures. The onset of quark matter at zero temperature is at $\SI{3}{\nsat}$ and shifts to lower densities for higher temperatures. As in the chiral broken phase, the isentropes initially have a very steep slope in the hadronic phase until they reach the first‑order phase boundary between the hadronic and 2SC phase. Once again the trajectories tend to follow this boundary, dropping sharply in temperature with increasing baryochemical potential along it to conserve entropy. 


Because we move along lines of constant entropy per baryon $s/n$, the
direction of this drop encodes which of the two coexisting phases carries the
larger $s/n$. 
The drop in temperature is thus an observation that
tells us how $s/n$ is ordered across the transition.
Whether this ordering can also be understood microscopically depends on the
transition. It can be justified only for the 2SC--CFL transition, where both
phases are described within the same NJL model. For the hadronic--2SC boundary no such
comparison is available, since the hadronic (DD2) and quark (NJL) phases
belong to different effective descriptions with no common quasiparticle basis;
there the larger entropy per baryon of the 2SC phase is simply what we read
off from the direction of the drop, not something that can be argued from the
degrees of freedom.

With this setup, mass-radius curves are calculated and shown in section \cref{sec:mr_chapter_transparent}. Matter in the center of the corresponding maximum-mass configuration is already shown in \cref{fig:pd_DD2NJL1} as filled circles on the trajectories. They indicate up to which maximum mass stable neutron star configurations can exist for this particular equation of state. It is observable that all proto-neutron star configurations studied here can exhibit color-superconducting phases
above a certain central energy density. The actual evolution trajectory will be explored in section~\cref{sec:mr_chapter_combined} by connecting the different proto-neutron star evolutionary stages with isolines for a conserved baryon number.

 An interesting feature and a consequence of the isentropic trajectory along the phase boundary is that there are neutron-star configurations that have outer layers that are hotter than the core, which can be concluded from the drop in temperature at the phase boundary. Such inverse temperature gradients may have an impact on thermal relaxation and neutrino transport inside PNSs. Exploring this would require implementing our hybrid equation of state in dynamical proto-neutron star simulations with neutrino transport \cite{Buras:2005rp, Roberts:2016lzn, Kuroda:2018gqq, Burrows:2019zce}.

\begin{figure}[t]
\includegraphics[width=0.5\textwidth]{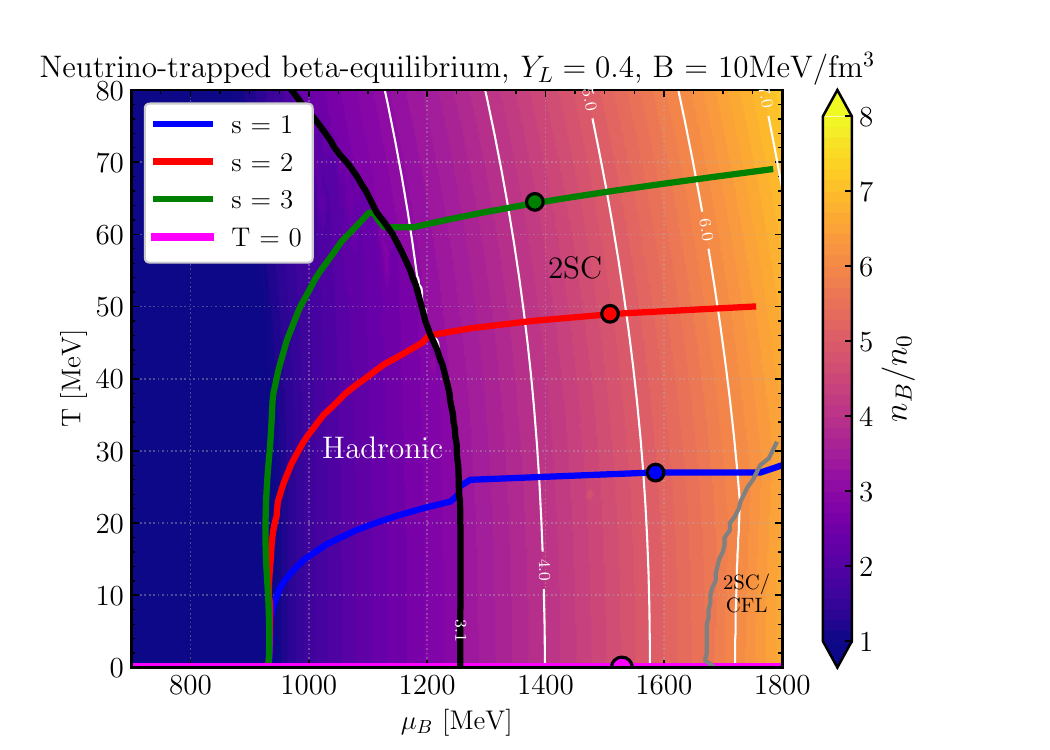}
\caption{Same as \cref{fig:pd_DD2NJL1}, but in neutrino-trapped beta-equilibrium, assuming an electron lepton fraction of $Y_L=0.4$. The hadronic, 2SC, and CFL phases are shown together with isentropic trajectories for different values of entropy-per-baryon $s=1,2,3$ and the zero-temperature limit. The color map indicates the baryon number density with isodensity contours at integer multiples of the nuclear saturation density.The filled circles indicate matter in the center of the maximum-mass configuration of the hybrid EoS, see the corresponding mass-radius curves in \cref{fig:mr_yl04}.}
\label{fig:pd_nobeta}
\end{figure}

In \cref{fig:pd_nobeta}, the phase diagram of the hybrid equation of state including neutrino-trapping is shown. The lepton fraction is fixed to a value of $Y_L = 0.4$ to get a picture of proto-neutron star conditions directly after the explosion. Compared to the case without any net neutrino number, there is no region in the phase diagram with a pure CFL phase anymore. As the CFL phase is charge neutral by definition additional leptons like electrons disfavor an onset at low densities. This feature is well known and has been observed in previous work, see \cite{Ruester:2005ib, Sandin:2007zr}. Instead, there appears a mixture of a 2SC and a CFL phase at low temperatures and high densities, which is for our parameter choices and the way of constructing the mixed phase the energetically more favorable state at these temperatures and densities. However, this mixed phase does not affect the results for the proto-neutron star at all, as the corresponding region of the phase diagram is not reached for stable configurations.

Imposing $Y_L = 0.4$ on the system leads to a shift of the hadronic isentropes to higher temperatures. With increasing lepton fraction, the electron and neutrino densities increase. This modifies the beta-equilibrium and charge neutrality conditions, which leads to a higher proton fraction and ensures that the numbers of protons and neutrons converge. In generel, equal amounts of protons and neutrons are much more favored which leads to an increase in the symmetry energy. As a result the equation of state and the entropy per baryon undergo corresponding modifications. This does not apply to the 2SC phase which is considerably less sensitive to isospin asymmetry than hadronic matter. Because of the pairing between up and down quarks it naturally favors symmetric matter, even in the absence of a high lepton fraction.

In summary, we find that all maximum NS configurations lie in the 2SC phase for the neutrino-trapped case, which leaves room for a possible CSC region in the cold NS to be explored in more detail in the following chapter. 

\subsection{Mass-radius curves of neutrino-transparent isentropic hybrid-star sequences}\label{sec:mr_chapter_transparent}

\begin{figure}[t]
\includegraphics[width=0.5\textwidth]{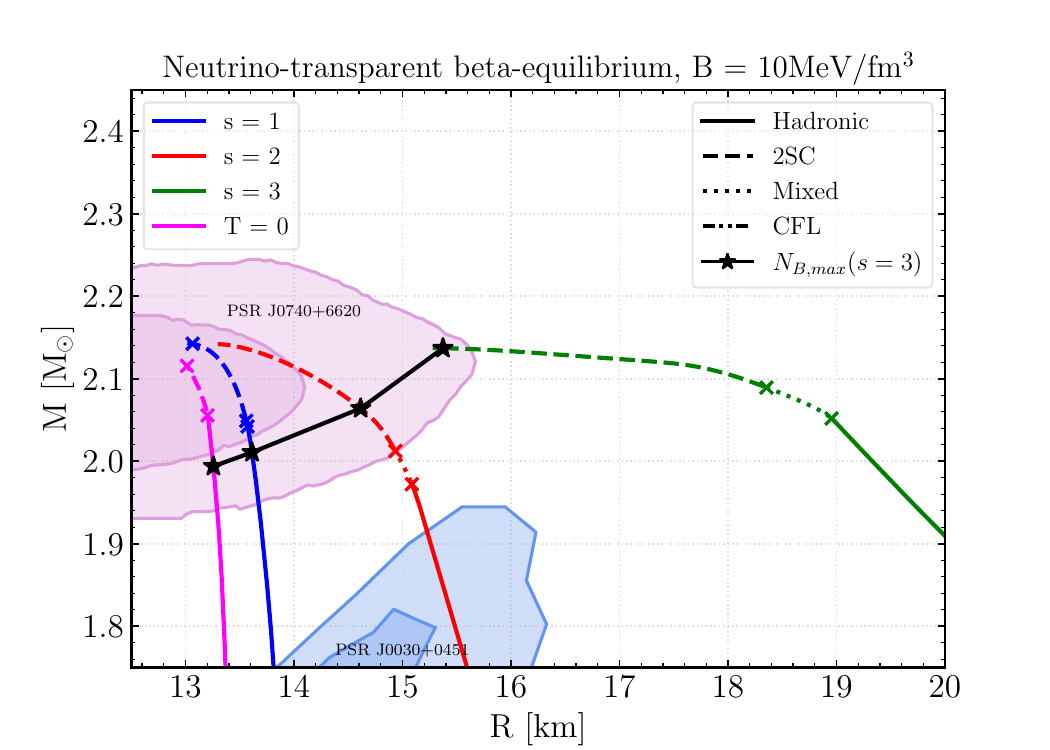}
\caption{Mass--radius relations for the DD2+NJL1 hybrid EoS under neutrino-transparent ($Y_\nu = 0$) beta equilibrium at constant entropy per baryon $s = 1,\,2,\,3$  and at $T = 0$. Line styles distinguish the phase at the stellar center: solid (hadronic), dashed (2SC), dotted (mixed hadronic--2SC, 2SC--CFL), and dash-dotted (CFL). Crosses mark the transition between two phases. Black lines show the isolines of constant baryon number ($N_B = \mathrm{const}$) indicating the evolutionary track by connecting configurations of equal baryon content from $s=3$ to $T=0$ states. Shaded regions show the 68\% and 95\% credibility contours for PSR~J0740+6620~\cite{Miller:2021qha, Riley:2021pdl} and
 PSR~J0030+0451~\cite{Miller:2019cac, Riley:2019yda}.}
\label{fig:mr_yn0}
\end{figure}

 We first investigate the mass-radius relation of isentropic hybrid-star sequences in neutrino-transparent matter, corresponding to the late stages of proto-neutron-star evolution and cold neutron stars. \cref{fig:mr_yn0,fig:mr_yl04} present the mass--radius sequences for the DD2+NJL1 hybrid EoS. \cref{fig:mr_yn0} shows the neutrino-transparent \d($Y_\nu = 0$) beta equilibrium case, while \cref{fig:mr_yl04} shows the neutrino-trapped matter with fixed lepton fraction $Y_L = 0.4$, typical of the early proto-neutron star (PNS) phase. Both figures display profiles at constant entropy per baryon $s = 1,\,2,\,3$ alongside the cold, $T = 0$ EoS.
The line style encodes the phase at the stellar center: solid (hadronic), dashed (2SC), dotted (mixed 2SC--CFL), and dash-dot-dot (CFL), with crosses marking the transition at different phases.
 
In the neutrino-transparent case (\cref{fig:mr_yn0}), all sequences exceed $2\,M_\odot$ and are consistent with the NICER constraint on PSR~J0740+6620~\cite{Fonseca:2021wxt,Miller:2021qha,Riley:2021pdl} (except $s=3$), with increasing entropy systematically expanding the radius and raising the maximum mass through thermal pressure support~\cite{Mariani:2016pcx, Hempel:2011mk}.

For the cold $T=0$ sequence, the stellar core remains purely hadronic up to the deconfinement onset density, then undergoes a direct transition from hadronic matter to a 2SC phase, and reaches a maximum-mass configuration featuring a small CFL core. This trend of the mass-radius curve becoming unstable after transitioning to a CFL configuration has been observed as well in \cite{Sandin:2007zr}. 

The finite-entropy sequences trace distinct trajectories across the $T$--$\mu_B$ phase diagram, causing the internal stellar phases to evolve with central density along individual constant-entropy paths. For the $s=1$ sequence—which provides the least thermal pressure support among our finite-entropy profiles—the stellar core is compressed to the highest central densities at its stability limit. This high central density allows the central core to transition sequentially from hadronic matter through a hadronic--2SC mixed phase, into a pure 2SC phase, and finally into a narrow 2SC--CFL mixed region at the maximum-mass configuration. As entropy increases to $s=2$ and $s=3$, higher thermal pressure and temperature suppress three-flavor pairing in the core. At $s=3$, the core transitions from the hadronic phase into the hadronic--2SC mixed phase, followed by a wide, prominent pure 2SC segment that persists all the way up to the maximum-mass configuration. Consequently, for the $s=3$ sequence, the lower maximum central density restricts the core to the pure 2SC branch over a wide range of radii, preventing it from reaching the CFL threshold before gravitational instability sets in. This thermal stabilization of the 2SC phase at higher temperatures has been observed in broader hybrid-star contexts~\cite{Blaschke:2010vd, Roark:2018uls}. 

Along the high-entropy sequences in \cref{fig:mr_yn0}, the mass-radius curves exhibit a characteristic behavior where the total stellar radius decreases as a function of increasing central density while the gravitational mass remains nearly constant. This flat structural plateau, appearing as a horizontal shift to the left in the mass–radius diagram, is a direct consequence of the phase transitions. The onset of color superconductivity softens the equation of state, causing the star to compactify as central density increases without sustaining a corresponding rise in gravitational mass. The microphysical softening associated with the phase transition fundamentally reduces the star's internal pressure support. Specifically, as the local temperature drops along the isentrope within the mixed phase, this loss of thermal pressure causes the equation of state to soften. Consequently, the star undergoes a stable structural compactification, characterized by a decreasing radius at a nearly invariant gravitational mass, which produces the characteristic flat horizontal plateau in the MR plane. However, as gravitational compression drives the central density higher, the core transitions into the pure 2SC phase, whereupon the equation of state re-stiffens. This high-density stiffening causes the gravitational mass to turn upward again on a secondary stable branch. 

It is precisely this two-step thermodynamic sequence: a transient thermal softening succeeded by a robust, high-density stiffening, that drives the onset of thermal twins, wherein macroscopically distinct stellar configurations sustain identical gravitational masses despite possessing disparate radii and internal structures~\cite{Hempel:2015vlg, Carlomagno:2023nrc}. Thermodynamically, this structural degeneracy is realized only when the derivative of the temperature with respect to the baryon density along the phase boundary is negative,
\begin{equation}
\frac{\mathrm{d}T}{\mathrm{d}n_B} < 0,
\end{equation}
This negative slope is a defining feature of a non-congruent phase transition \cite{Roark:2018uls}.

As the isentropes enter the hadronic--2SC mixed phase, the temperature decreases as the baryon chemical potential increases along the isentrope, see \cref{fig:pd_DD2NJL1}. The large heat capacity of the mixed phase, where newly liberated quark degrees of freedom absorb entropy at nearly fixed temperature, drives this behavior. For $s = 2$, the temperature drops from around $T=55\,\mathrm{MeV}$ at the mixed-phase onset to a plateau at $T = 48\,\mathrm{MeV}$ throughout the pure 2SC region; this nearly isothermal segment is the broad dashed portion in the M-R diagram. For $s = 3$, the temperature declines from $T=72\,$MeV to $T=63\,$MeV. 
 This localized thermal pressure drop and the abundance of a mixed phase soften the total EoS, causing the star to compactify into a thermal twin configuration: distinct stellar configurations sustaining identical masses despite differing radii and internal structures. Past studies have observed these identical macroscopic plateaus, which confirm that non-congruent deconfinement transitions in hot proto-neutron stars produce these thermal twins~\cite{Roark:2018uls, Hempel:2015eoj, Hempel:2015vlg, Carlomagno:2023nrc}.

The constant-baryon-number ($N_B = 2.74 \times 10^{57}$) tracks illustrate this thermal evolution. Tracing the path originating at the $s = 3$ maximum mass, the star begins with a pure 2SC core. As it cools to the $s = 2$ sequence, the core remains within the pure 2SC domain. Upon cooling further to $s = 1$, the loss of thermal pressure support causes the track to shift into the purely hadronic phase. This track remains entirely within the hadronic branch all the way down to its final, cold $T = 0$ configuration. This shifting core composition along a single $N_B$ path highlights how rapid thermal evolution can temporarily revert a hybrid star's core to hadronic matter before it achieves its final, cold dense state~\cite{Pons:1998mm, Hempel:2015eoj}.

\subsection{Mass-radius curves of isentropic hybrid-star sequences with neutrino trapping}\label{sec:mr_chapter_trapped}

\begin{figure}[t]
\includegraphics[width=0.5\textwidth]{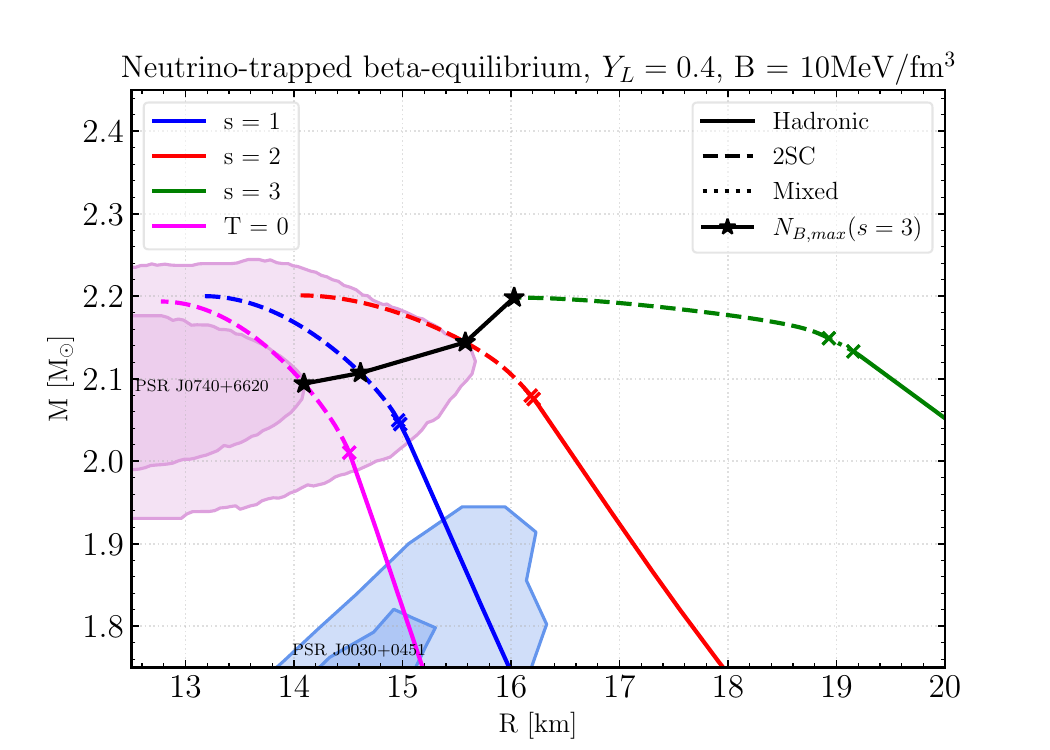}
\caption{Same as \cref{fig:mr_yn0} but for neutrino-trapped matter with fixed total lepton fraction $Y_L = 0.4$, characteristic of the early PNS phase. Compared to the $Y_\nu = 0$ case shown in \cref{fig:mr_yn0}, the radii are larger and the isolines of constant baryon number are at higher gravitational masses, so that the endpoints of the isolines at $T=0$ are all within the 2SC phase.}
\label{fig:mr_yl04}
\end{figure}

We now consider the impact of neutrino trapping on the mass-radius relation of hybrid-star sequences, which is relevant for the early proto-neutron-star evolution.
Under neutrino trapping conditions ($Y_L = 0.4$, \cref{fig:mr_yl04}), the structural composition of the sequences shifts due to the preservation of local lepton number, which modifies the equilibrium composition and equation of state. At this high lepton fraction, the hadronic--2SC phase boundary is displaced to lower baryon chemical potential across all relevant temperatures. The 2SC--CFL phase transition is pushed entirely beyond the density range of stable stellar configurations. As a result, the finite-entropy profiles present a simplified phase layout compared to the neutrino-transparent case: the stars remain purely hadronic over an extended radius range, transition into a hadronic--2SC mixed phase, and terminate at their maximum-mass endpoints possessing exclusively pure 2SC cores. The three-flavor pairing phases (mixed 2SC--CFL and pure CFL) are entirely absent from these curves, as confirmed by the $Y_L = 0.4$ phase diagram where the maximum-mass endpoints terminate well within the 2SC domain.


The macroscopic MR curves directly reflect these microphysical shifts. This behavior exhibits a strong dependency on the entropy of the sequence, driven by the structural alterations of the phase boundaries under neutrino trapping. 

In general, under neutrino-trapping, the radius of the configurations becomes larger due to the additional pressure support of the leptons. The maximum masses experience only a minor increase. Another interesting feauture is the behavior of the isoline in both cases. The decrease in mass from the $s = 3$ to the $T = 0$ configuration is larger in the neutrino-transparent case, which indicates a larger difference between the gravitational and baryonic mass for $T = 0$ in a neutrino-transparent configuration than in a neutrino-trapped configuration. 
Furthermore, 
the high-entropy curves ($s = 2, 3$) still maintain a mild flat plateau in their mass-radius profiles as they enter into the 2SC phase.
However, the thermal twin behavior is less pronounced here than in the neutrino-transparent case, as the restrictions imposed by lepton conservation temper the softening typically induced by unconstrained deconfinement transitions at finite temperature~\cite{Roark:2018uls, Hempel:2015eoj}.

The constant-baryon-number ($N_B = 2.79 \times 10^{57}$) tracks under neutrino-trapped conditions ($Y_L = 0.4$) illustrate the early structural cooling of the PNS. Tracing the path originating at the $s = 3$ maximum mass, the core begins in the pure 2SC phase. As the star cools sequentially through the $s = 2$ and $s = 1$ isentropes down to the cold $T = 0$ configuration, the track remains entirely within the pure 2SC domain. This continuous stability shows that under neutrino trapping, the 2SC core is robust against thermal pressure loss and does not revert to hadronic or mixed phases during early cooling~\cite{Pons:1998mm, Roark:2018uls}.

\begin{figure}[t]
\includegraphics[width=0.5\textwidth]{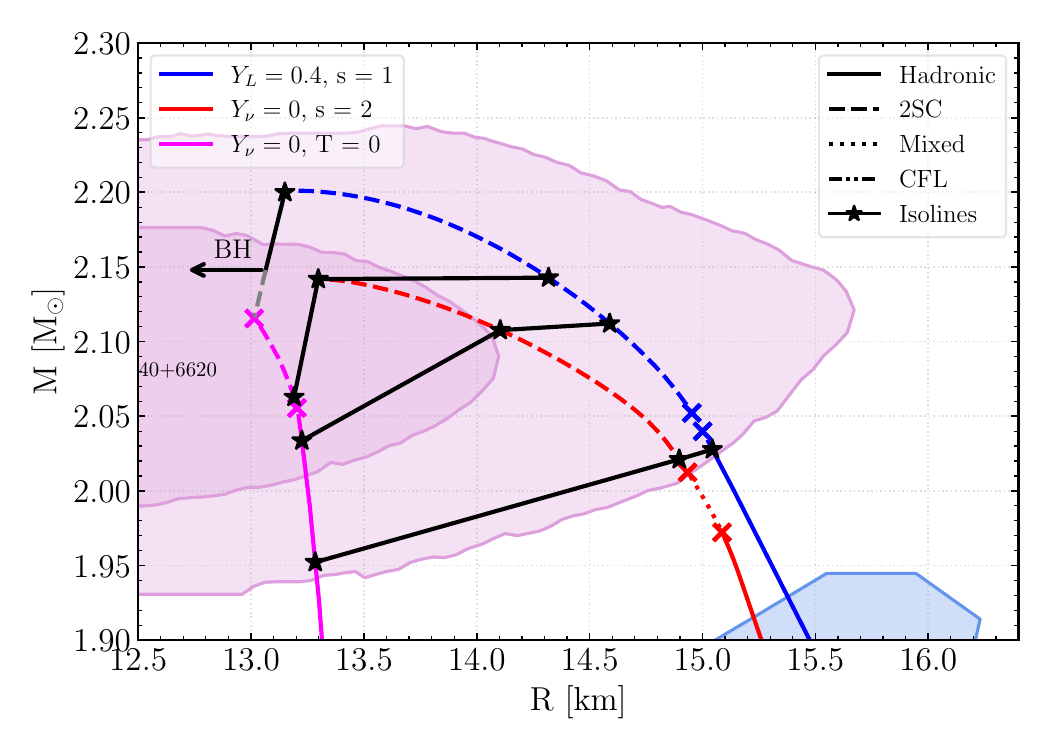}
\caption{Zoomed mass--radius diagram in the high-mass region comparing the $Y_L = 0.4$, $s = 1$, and $Y_\nu = 0$, $s = 2$ isentropic sequences together with the cold $T = 0$ sequence for the DD2+NJL1 EoS. Black lines are constant-$N_B$ evolutionary tracks connecting the birth state ($Y_L = 0.4$, $s = 1$) through the post-deleptonization state ($Y_\nu = 0$, $s = 2$) to the cold remnant ($T = 0$). Line styles and observational constraints are the same as in \cref{fig:mr_yn0}.}
\label{fig:mr_combined}
\end{figure}

\subsection{PNS evolution in the mass-radius diagram}\label{sec:mr_chapter_combined}

Having established the structural properties of hybrid-star sequences in both neutrino-transparent and neutrino-trapped matter, we now follow the evolution of proto-neutron stars along trajectories of fixed baryon number.
\cref{fig:mr_combined} tracks the thermodynamic evolution of the remnant by overlaying constant-baryon-number ($N_B = \text{const.}$) isolines (black starred tracks) across three representative astrophysical stages: the early PNS birth state ($Y_L = 0.4$, $s=1$), the post-deleptonization stage ($Y_\nu = 0$, $s=2$), and the final cold isolated remnant ($Y_\nu = 0$, $T=0$). Depending on the initial baryon mass bound at birth, the structural softening or stiffening induced by the phase transitions splits the stellar configurations which include color-superconducting matter into four distinct evolutionary scenarios:

\begin{enumerate}
\item \textbf{Delayed gravitational collapse to a black hole:} This scenario applies to PNSs born with a high baryon mass, such as the topmost isoline originating at the $Y_L = 0.4$ maximum-mass configuration ($M \approx 2.20\,M_\odot$, $R \approx 13.22$~km). As the star cools and deleptonizes, it sheds the crucial lepton and thermal pressure support required to balance its immense gravity. Because this initial baryon number exceeds the maximum stable mass of the subsequent $Y_\nu = 0, s = 2$ phase, the star becomes unstable and collapses into a black hole midway through its evolution, before it can ever reach the cold $T = 0$ sequence. The mathematical extension of this track to the $T=0$ sequence is shown as a gray dashed line in \cref{fig:mr_combined} strictly for illustrative completeness. 

\item \textbf{2SC $\rightarrow$ 2SC $\rightarrow$ 2SC sequence:} This path describes stable high-mass remnants, exemplified by the second isoline from the top, which originates on the dashed branch of the birth state ($M \approx 2.15\,M_\odot$) with a pure 2SC core. Upon deleptonization to the $Y_\nu = 0$, $s=2$ sequence, the track remains stable within the 2SC phase, passing through its respective maximum-mass configuration. During final cooling to $T=0$, the compression shifts the central state toward higher densities, allowing the core to stay in a stable 2SC configuration.

\item \textbf{2SC $\rightarrow$ 2SC $\rightarrow$ Hadronic sequence:} This track corresponds to mass configurations of ${M \approx 2.05\,M_\odot - 2.1\,M_\odot}$. The star is born with a sufficiently high central density to possess a pure 2SC core on the $Y_L = 0.4$ curve. During the deleptonization phase to the $s=2$ sequence, the star contracts and its core remains within the pure 2SC phase. However, during final cooling to $T=0$, the loss of thermal pressure support causes the central density to fall back below the cold quark-deconfinement threshold. The track crosses the phase boundary and terminates on the cold hadronic branch, effectively reverting a core that was once quark matter back into purely hadronic matter. Although we do not show it in \cref{fig:mr_combined} there is also the case of a core that starts of as a mixture of 2SC and hadronic matter. It will first transform into pure 2SC matter after deleptonization ($Y_\nu = 0$, $s = 2$), and then to hadronic matter after cooling ($T=0$).

\item \textbf{Hadronic $\rightarrow$ 2SC $\rightarrow$ Hadronic sequence:} For remnants with masses of $M \lesssim 2.0\,M_\odot$, represented by the bottommost isolines, the star is born on the solid, purely hadronic branch of the $Y_L=0.4$, $s=1$ curve. As the star sheds its trapped neutrinos and transitions to the $Y_\nu=0$, $s=2$ sequence, the star's increased temperature allows for a shift of the core across the phase boundary into a pure 2SC (case shown in \cref{fig:mr_combined}) or a hadronic-2SC mixed state. Upon final cooling to $T=0$, the path tracks back into the hadronic domain in the core.

\end{enumerate}

These four distinct structural sequences give rise to a heavily non-monotonic radius evolution along individual evolutionary tracks. The star undergoes an initial rapid contraction as trapped neutrinos escape during the early deleptonization phase ($Y_L = 0.4 \rightarrow Y_\nu = 0$), followed again by a notable contraction during the subsequent thermal cooling stage. This unique macroscopic signature fundamentally differentiates hybrid models from purely hadronic core-collapse simulations~\cite{Pons:1998mm, Prakash:1996xs}. From the fact that there is no 2SC $\rightarrow$ Hadronic $\rightarrow$ Hadronic sequence it follows that 2SC quark matter survives deleptonization and that the conversion to hadronic matter only happens in the subsequent cooling. We furthermore note that the number of different evolutionary, color superconductivity including paths, may change if one switches to a different model or chooses different parameters.

\section{Summary}
In this work we studied the influence of color-superconducting matter on the compositional evolution of a proto-neutron star. The color-superconducting matter was modeled within the RG-consistent NJL model, which allows thermodynamically consistent calculations at high chemical potentials without running into regularization artifacts. 
In particular, we focused on the phase transitions occuring in the stellar core under the constraint of constant baryon number conservation following the evolutionary trajectory of a proto-neutron star. 
The cases of neutrino-transparent and neutrino-trapped beta equilibrium were studied separately in order to clearly distinguish the impact of temperature and a fixed lepton fraction. 

As isentropes are constrained to follow the hadron-quark phase boundary in the $T$--$\mu_B$ plane, all isentropic PNS sequences studied include such a mixed phase. Consequently, if the temperature decreases with increasing density along an isentrope in the mixed phase, there are stars on the sequence for which the temperature of the 2SC quark core is less than a layer of less-dense surrounding hadronic or hadronic-2SC mixed matter.

The plausibility of such a temperature profile in nature requires further investigation using more detailed dynamical calculations that include neutrino transport, convection, and thermal diffusion.

We do not obtain PNS stellar sequences with pure CFL matter in the core. In particular, the CFL phase is absent in the stable part of those stellar sequences which assume the neutrino-trapped condition of the new-born star ($Y_L=0.4$). This is due to the general fact that the CFL phase is an electronic insulator, such that it gets disfavored against the 2SC phase by the addition of electrons to the system. In the neutrino-transparent case, the maximum mass of the isentropic stellar sequences lies in the 2SC phase (except for $s=1$, for which it lies in a 2SC-CFL mixed phase). This is due to the number of effective degrees of freedom being smaller in the CFL (all quarks are paired) than in the 2SC phase (unpaired blue quarks, unpaired strange quarks of all colors), thus leading to the 2SC-CFL phase boundary bending towards increasing $\mu_B$ with increasing $T$.


We calculated mass--radius diagrams along fixed isentropes for both the neutrino-transparent ($Y_\nu = 0$) and neutrino-trapped ($Y_L = 0.4$) configurations. By tracing constant-baryon-number ($N_B$) evolutionary trajectories starting from the maximum-mass configurations of the hot $s = 3$ isentropes down to the cold $T = 0$ configuration, we reveal a clear difference in the final stellar core compositions. In the neutrino-transparent case, the loss of thermal pressure support causes the evolutionary track of the maximum-mass to shift entirely into the hadronic branch, meaning the resulting cold remnant contains only hadronic matter at its center. Conversely, under neutrino-trapped conditions, the enhanced stability of the 2SC phase prevents this reversion, making it entirely possible to retain a pure 2SC core in the final cold configuration.

By mapping the full multidimensional transition from a hot, neutrino-trapped birth state to a cold, neutrino-transparent final state, we identify four distinct evolutionary scenarios for the core composition. 
With decreasing mass of the initial neutrino-trapped PNS configuration, we identify four different evolution paths in our model:
\begin{enumerate}
    \item 2SC $\rightarrow$ Collapse to black hole,
    \item 2SC $\rightarrow$ 2SC $\rightarrow$ 2SC,
    \item 2SC $\rightarrow$ 2SC $\rightarrow$ Hadronic, and
    \item Hadronic $\rightarrow$ 2SC $\rightarrow$ Hadronic.
\end{enumerate}
The final $T=0$ stellar configurations are fully compatible with the astrophysical constraints from PSR~J0740+6620.

Our macrostructural results are tightly bound to the microphysical structure of the underlying $T$--$\mu_B$ phase diagrams. The temperature and lepton-fraction dependence of the phase boundaries determines when and where color-superconducting phases emerge during proto-neutron-star evolution. As the remnant deleptonizes and cools, the shifting phase boundaries alter the active degrees of freedom, causing these quark phases in the beginning of evolution to either transform or vanish entirely in the cold, isolated NS configurations.

In conclusion, our findings demonstrate that CSC matter dictates the early, dynamical evolution of dense stellar remnants, whereas stable, cold neutron stars are predominantly governed by hadronic physics. The emergence of the color-superconducting phases investigated in this work may eventually be probed through multimessenger observations from core-collapse supernovae and neutron star mergers. Megahertz gravitational waves from neutron star mergers have already been identified as promising indicators for the appearance of a first-order phase transition at high densities \cite{Blas:2022xco}, and can also be used to study phase transitions during proto-neutron star evolution \cite{Bleau:2026ala}. In the four scenarios discussed in this work, one would even expect two signatures from two phase transitions (Hadronic $\rightarrow$ 2SC $\rightarrow$ Hadronic) giving rise to the possible delayed (dis-)appearance of two different MHz gravitational-wave signal forms. It is possible that there are two phase transitions in proto-neutron star evolution originating from the appearance of the 2SC and the CFL phase even for the neutrino-transparent stage, see also \cref{sec:app} for such cases. 

The measurement of the supernova neutrino signal could deliver additional information on the appearance of CSC in the newly born proto-neutron stars. Since hadronic matter, 2SC matter, and CFL matter all differ in transport properties such as neutrino emissivities and cooling behavior, it could be possible to observe these characteristics in the neutrino signal \cite{Carter:2000xf}. We stress that in three of our scenarios (scenario 1, 3 and 4) it might well be that CSC phases are present only in the fleeting moments of hot proto-neutron stars right after a core-collapse supernova. CSC phases could also appear in the short moment of heated neutron-star matter during a neutron star merger as the conditions for beta-equilibrium are the same. 

Future advances in gravitational-wave and neutrino astronomy may therefore provide a unique opportunity and a diagnostic tool to test for the existence of these exotic CSC phases and to probe the behavior of strongly interacting matter under extreme conditions.


\section*{Acknowledgement}
We thank Mark G. Alford for valuable discussions, insightful explanations, and continuous advice throughout this project. We are also grateful to Michael Buballa for helpful discussions and for several valuable suggestions.
The authors acknowledge support by the Deutsche Forschungsgemeinschaft (DFG, German Research Foundation) through the CRC-TR 211 'Strong-interaction matter under extreme conditions' -- project number 315477589 -- TRR 211. I. A. R. gratefully acknowledges support from the Deutsche Forschungsgemeinschaft (DFG, German Research Foundation) – Project Number 579861443 and also in part by the Alexander von Humboldt Foundation through a Humboldt Research Fellowship. M. H. is supported by the GSI F\&E.

\appendix
\section{Phase diagrams with different bag pressure}\label{sec:app}

\begin{figure}[t]
\includegraphics[width=0.5\textwidth]{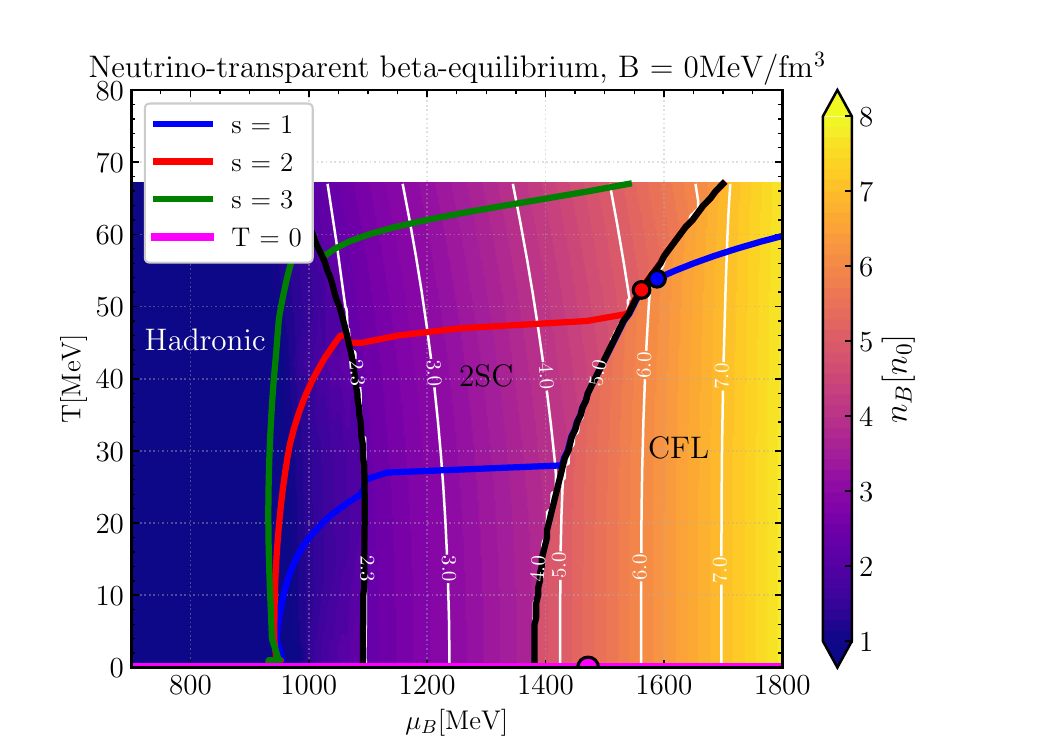}
\caption{Same as \cref{fig:pd_DD2NJL1}, but with bag constant set to zero, $B$ = $\SI{0}{\MeV\per\fm^3}$.}
\label{fig:pd_NJL1_B0}
\end{figure}

\begin{figure}[t]
\includegraphics[width=0.5\textwidth]{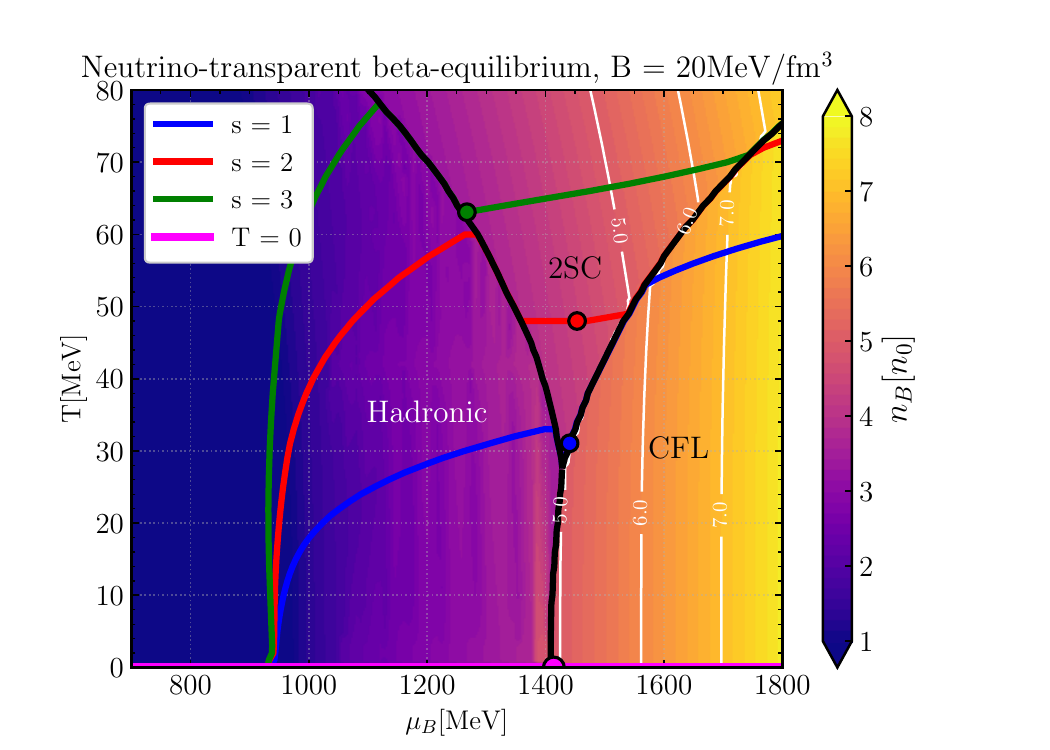}
\caption{Same as \cref{fig:pd_DD2NJL1}, but with bag constant, $B$ = $\SI{20}{\MeV\per\fm^3}$.}
\label{fig:pd_NJL1_B20}
\end{figure}

\cref{fig:pd_NJL1_B0,fig:pd_NJL1_B20} show phase diagrams of the RG-NJL model matched to the DD2 EoS with a bag pressure of $\SI{0}{\MeV\per\fm^3}$ and $\SI{20}{\MeV\per\fm^3}$, respectively. A  vanishing bag constant ($B= \SI{0}{\MeV\per\fm^3}$) is excluded from our primary analysis because it fails to produce a stable thermodynamic matching point with the hadronic EoS at higher temperatures. Conversely, adopting a higher bag constant of $B = \SI{20}{\MeV\per\fm^3}$ entirely suppresses the 2SC phase at zero temperature,  resulting in a direct transition from hadronic matter to the CFL phase. Because this transition is pushed to a significantly higher baryon density of around $5n_0$, color-superconducting cores would only manifest in the most massive neutron star configurations, if they appear at all. Interestingly, this high-bag-constant scenario generates a triple point where the hadronic, 2SC, and CFL phases coexist, a region that is explicitly traversed by the $s=1$ isentrope.


\bibliographystyle{apsrev4-112}
\bibliography{bib} 

\end{document}